\documentclass[11pt,a4paper,reqno]{article}
\usepackage{mysty}
\usepackage{fullpage}

\usepackage[thinlines]{easytable}

\SetLabelAlign{parright}{\parbox[t]{\labelwidth}{\raggedleft{#1}}}
\setlist[description]{style=multiline,topsep=4pt,align=parright}

\makeatletter
\let\reftagform@=\tagform@
\def\tagform@#1{\maketag@@@{(\ignorespaces\textcolor{black}{#1}\unskip\@@italiccorr)}}
\newcommand{\iref}[1]{\textup{\reftagform@{\tcr{\ref{#1}}}}}
\makeatother

\newenvironment{affiliations}{%
    \setcounter{enumi}{1}%
    \setlength{\parindent}{0in}%
    \slshape\sloppy%
    \begin{list}{\upshape$^{\arabic{enumi}}$}{%
        \usecounter{enumi}%
        \setlength{\leftmargin}{0in}%
        \setlength{\topsep}{0in}%
        \setlength{\labelsep}{0in}%
        \setlength{\labelwidth}{0in}%
        \setlength{\listparindent}{0in}%
        \setlength{\itemsep}{0ex}%
        \setlength{\parsep}{0in}%
        }
    }{\end{list}\par\vspace{12pt}}

\begin{document}

\title{Digital System Reconstruction by Pairwise Transfer Entropy}
\author{Zhong-Qi Kyle Tian\textsuperscript{1}, Douglas Zhou\textsuperscript{1,\footnote{zdz@sjtu.edu.cn}}, David Cai\textsuperscript{1,2,3}}

\date{}
\maketitle
\begin{affiliations}
 \item School of Mathematical Sciences, MOE-LSC, and Institute of Natural Sciences, Shanghai Jiao Tong University, Shanghai, China	
 \item Courant Institute of Mathematical Sciences and Center for Neural Science, New York University, New York, NY, United States of America
 \item NYUAD Institute, New York University Abu Dhabi, Abu Dhabi, United Arab Emirates\\
\end{affiliations}

\begin{abstract}
Transfer entropy (TE) is an attractive model-free method to detect
causality and infer structural connectivity of general digital systems.
However it relies on high dimensions used in its definition to clearly
remove the memory effect and distinguish the direct causality from
the indirect ones which makes it almost inoperable in practice. In
this work, we try to use a low order and pairwise TE framework with
binary data suitably filtered from the recorded signals to avoid the
high dimensional problem. Under this setting, we find and explain
that the TE values from the connected and unconnected pairs have a
significant difference of magnitude, which can be easily classified
by cluster methods. This phenomenon widely and robustly holds over
a wide range of systems and dynamical regimes. In addition, we find
the TE value is quadratically related to the coupling strength and
thus we can establish a quantitative mapping between the causal and
structural connectivity. 
\end{abstract}

\section*{Introduction}
The structural connectivity of networks is of great importance to
understand their function in many scientific fields like the genetic
regulatory networks and neuronal circuitry networks. When only limited
knowledge of the underlying dynamics is given, it is almost impossible
to identify the actual connectivity directly in experiment, although
abundant time series data generated by individual nodes in networks
can be recorded much more easily in today's technology. Based on these
massive quantities of time series, one may use statistical approaches
to identify the causal connectivity and further infer the function
of the network \cite{strogatz2001exploring,guelzim2002topological,ding608035granger,szekely2014partial}.
Transfer entropy (TE), an information-theoretic measure, attracts
more and more interest due to its excellent model-free property \cite{schreiber2000measuring,honey2007network,gourevitch2007evaluating,vicente2011transfer},
requiring no detailed knowledge of the system. Unlike the Grange causality
(GC) analysis \cite{ding608035granger,guo2008uncovering,ge2009novel,Zhou2013},
which is generally limited to linear dynamics. For a system of two
nodes $X$ and $Y$ connected as $Y\rightarrow X$, the idea of TE
is that the history information of $Y$ is helpful to predict the
future of $X$, so the uncertainty about the future of $X$ conditioned
on its own memory is reduced by adding the past information of $Y$.
This reduction of uncertainty is quantified by TE. Because of its
conceptional simplicity and model-free property, TE has been widely
and successfully applied to causal problems in many fields, such as
neuron studies, social media and financial market \cite{vicente2011transfer,honey2007network,dimitrov2011information,ver2012information,li2013risk}. 

Despite the conceptual apeal of TE, it suffers the problem of ``the
curse of dimensionality'' \cite{indyk1998approximate,friedman1997bias,berchtold1998pyramid,runge2012escaping,rust1997using,hinneburg1999optimal}
for a successful application. For example, to clearly measure the
causality from $Y$ to $X$, the memory of $X$ should be well conditioned
\cite{wiener1956theory} which is often very long. Besides for large
networks, the intermediate nodes should also be taken into account
in case the computed causality is only indirect such as in the case
of $Y\rightarrow Z\rightarrow X$. Then the high dimensional problem
occurs. However the data recorded in experiments is limited, so TE
is often applied with some dimension reduction, $e.g.,$ take a bivariate
setting and truncate the length of memory (low orders in TE's definition)
\cite{schreiber2000measuring,gourevitch2007evaluating,vicente2011transfer,ito2011extending,runge2012escaping}
or make a specific prior assumption of the distribution of the signals,
such as the Gaussian distribution \cite{kraskov2004estimating,Moon1995,kaiser2002information,roudi2009pairwise,Steuer2002}.
To reduce dimension, we use low order and pairwise TE and use binary
time series filtered from the raw signals. Then a basic question is
that can the reduced TE framework successfully infer the causal connectivity.
Moreover the causal connectivity revealed by TE is only effective
or functional \cite{friston1994functional}, then how to infer the
structure connectivity from the TE causal connectivity? 

In this Letter, we consider three quite different and typical classes
of non-linear networks to show the wide application of TE: the Hodgkin-Huxley
(HH) neuronal network, Lorenz network and discrete Logistic map systems.
Our numerical results show that there is a significant difference
of order of magnitude between the connected and unconnected TE values.
Thus we can set a proper threshold to classify the TE values into
two groups where the larger ones are inferred from the connected pairs.
The inferred TE connectivity is highly consistent with the structure
connectivity over a wide range of dynamical regimes for all the three
systems. Our theoretical analysis show that the TE values are quadratically
related to the coupling strengths, thus we can establish a direct
map between these two types of connectivity. It is worthwhile pointing
out that we only use the spike timing of HH neurons to successfully
infer the structure connectivity of HH networks. The spikes of individual
neurons in a population can be easily recorded in experiment like
by the calcium imaging \cite{grewe2010high,stosiek2003vivo} and multi-electrode
array methods \cite{litke2004does,shimono2014functional}. Our work
may provide an operable and efficient TE framework to predict the
structural connectivity of these neuronal systems.

\section*{Results}
We first consider a Hodgkin-Huxley (HH) neuronal network with $N$
excitatory nodes (neurons). The HH model is widely used to simulate
neuronal networks for computational neural scientists \cite{abbott1990model,kistler1997reduction,pospischil2008minimal}.
The dynamics of the $i$th HH node is governed by 

\begin{equation}
\begin{cases}
\begin{aligned}C\frac{dV_{i}}{dt} & =-(V_{i}-V_{Na})G_{Na}m_{i}^{3}h_{i}-(V_{i}-V_{K})G_{K}n_{i}^{4}-(V_{i}-V_{L})G_{L}+I_{i}^{\textrm{input}}\\
\frac{dm_{i}}{dt} & =(1-m_{i})\alpha_{m}(V_{i})-m_{i}\beta_{m}(V_{i})\\
\frac{dhi}{dt} & =(1-h_{i})\alpha_{h}(V_{i})-h_{i}\beta_{h}(V_{i})\\
\frac{dn_{i}}{dt} & =(1-n_{i})\alpha_{n}(V_{i})-n_{i}\beta_{n}(V_{i})
\end{aligned}
\end{cases}\label{eq: HH}
\end{equation}
where $C$ is the membrane capacitance and $V_{i}$ is its membrane
potential, $m_{i}$, $h_{i}$ and $n_{i}$ are gating variables to
describe the sodium and potassium currents, $V_{Na},V_{K}$ and $V_{L}$
are the reversal potentials for the sodium, potassium and leak currents,
respectively, $G_{Na},$$G_{K}$ and $G_{L}$ are the corresponding
maximum conductances, $\alpha$ and $\beta$ are empirical functions
of $V$ \cite{hodgkin1952quantitative,dayan2001theoretical,gerstner2002spiking},

\begin{equation}
\begin{aligned} & \alpha_{m}(V_{i})=\frac{0.1V_{i}+4}{1-\exp(-0.1V_{i}-4)} &  & \beta_{m}(V_{i})=4\exp(-(V_{i}+65)/18)\\
& \alpha_{h}(V_{i})=0.07\exp(-(V_{i}+65)/20) &  & \beta_{h}(V_{i})=\frac{1}{1+\exp(-3.5-0.1V_{i})}\\
& \alpha_{n}(V_{i})=\frac{0.01V_{i}+0.55}{1-\exp(-0.1V_{i}-5.5)} &  & \beta_{n}(V_{i})=0.125\exp(-(V_{i}+65)/80)
\end{aligned}
\end{equation}

The input current $I_{i}^{\textrm{input}}$ is given by $I_{i}^{\textrm{input}}=-G_{i}(t)(V_{i}-V_{G})$
with

\begin{equation}
\frac{dG_{i}(t)}{dt}=-\frac{G_{i}(t)}{\sigma_{r}}+H_{i}(t)
\end{equation}

\begin{equation}
\frac{dH_{i}(t)}{dt}=-\frac{H_{i}(t)}{\sigma_{d}}+f\sum_{l}\delta(t-T_{il}^{F})+\sum_{j\neq i}A_{ij}\sum_{l}S\delta(t-T_{jl}^{S})\label{eq:f input}
\end{equation}
where $V_{G}$ is the reversal potential, $G_{i}(t)$ is the conductance,
$H_{i}(t)$ is an additional parameter to describe $G_{i}(t)$, $\sigma_{r}$
and $\sigma_{d}$ are fast rise and slow decay time scale, respectively,
and $\delta(\cdot)$ is the Dirac delta function. The second term
in Eq. (\ref{eq:f input}) is the feedforward input with magnitude
$f$ and the input time $T_{il}^{F}$ is generated from a Poisson
process with rate $\nu$. The last term in Eq. (\ref{eq:f input})
is the synaptic interactions in the network, where $\mathbf{A}=(A_{ij})$
is the adjacency matrix, $S$ is the coupling strength. When the voltage
of the $i$th node $V_{i}$, evolving continuously according to Eq.
(\ref{eq: HH}), reaches the threshold $V^{\textrm{th}}$, we say
it fires a spike at this time and denote it by $T_{il}^{S}$. Instantaneously,
all its postsynaptic nodes will receive this spike and their corresponding
parameter $H$ will jump by an amount of $S$. We record the spike
times of each node and transform them into binary time series by a
small time bin $\Delta t$ with the value 1 for a spike event in the
interval and 0 otherwise.

For two binary random processes $X,Y$ with states $x$ and $y$,
respectively, the TE \cite{schreiber2000measuring} from $Y$ to $X$
is defined by 

\begin{equation}
T_{Y\rightarrow X}(\tau)=\sum p(x_{n+1},x_{n}^{(k)},y_{n-\tau}^{(l)})\log\frac{p(x_{n+1}|x_{n}^{(k)},y_{n-\tau}^{(l)})}{p(x_{n+1}|x_{n}^{(k)})}\label{eq:TE}
\end{equation}
where $\tau$ is a proper time delay of interest, $x_{n}^{(k)}=(x_{n},x_{n-1},...,x_{n-k+1})$
and $y_{n-\tau}^{(l)}=(y_{n-\tau},y_{n-\tau-1},...,y_{n-\tau-l+1})$,
$k,l$ are the orders (memory) of $X$ and $Y$, respectively. According
to Wiener\textquoteright s principle \cite{wiener1956theory}, we
should use a sufficiently large value of $k$ to remove the memory
effect of $X$. Then we can call $Y$ is causal to $X$ if the information
of the past of $Y$ improves the prediction of $X$. However, it would
increase the dimension greatly and make it inoperable in practice. 

\begin{figure}[H]
	\begin{centering}
		\includegraphics[width=1\textwidth]{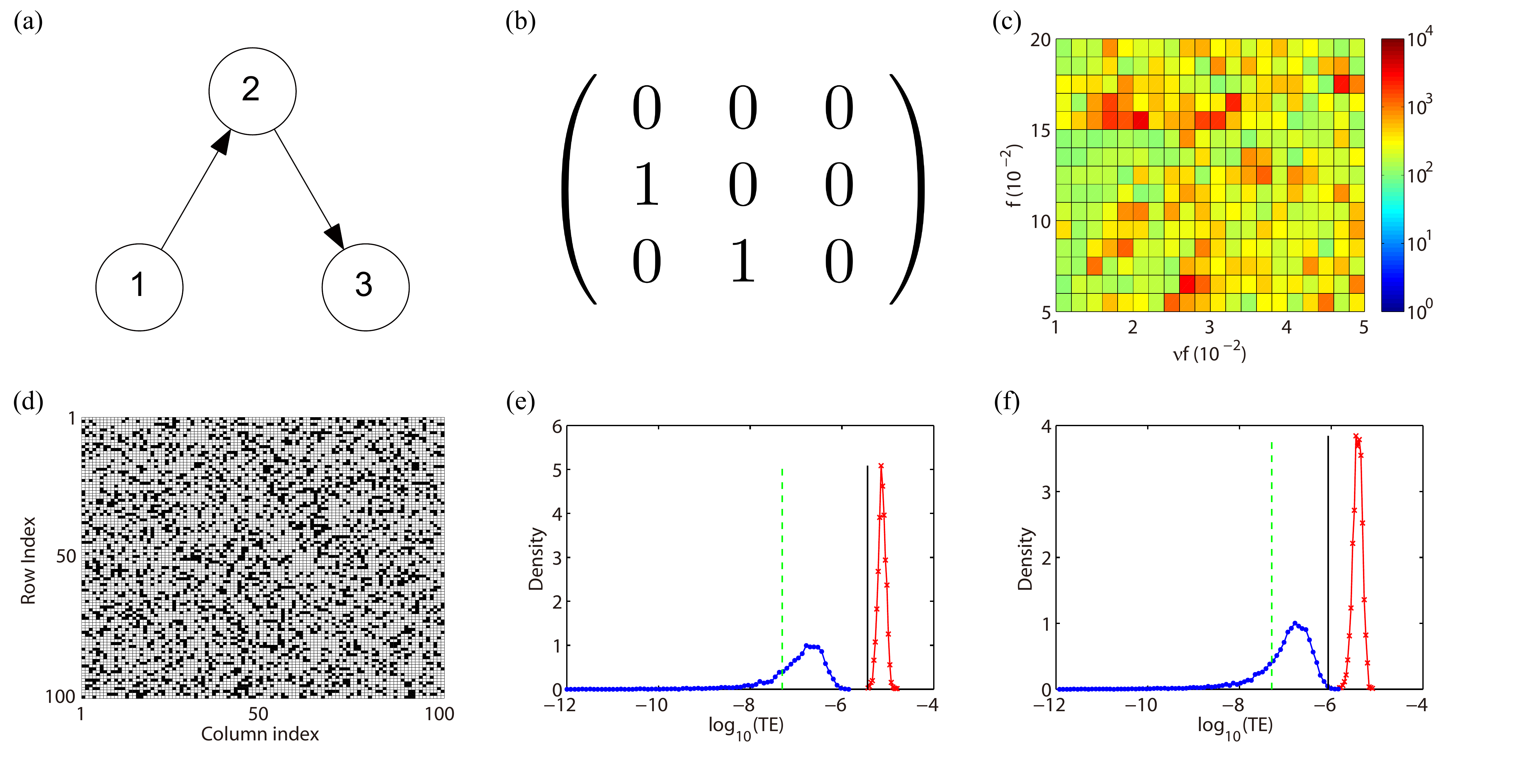}
		\par\end{centering}
	\caption{(a) A network with three nodes. (b) Predicted adjacency matrix constructed
		by TE by the difference of magnitude which captures the structure
		connectivity in (a). (c) The ratio of mean TE values from connected
		and unconnected pairs. (d) The adjacency matrix of a randomly connected
		network of 100 nodes with the black color indicating $A_{ij}=1$ and
		white color for $A_{ij}=0$. The element $A_{ij}$ has a probability
		of 25\% to be 1. (e) The density of $\log_{10}(\textrm{TE})$ values
		for the network in (d). (f) The density of $\log_{10}(\textrm{TE})$
		values after adding a noise generated from the uniform distribution
		$U(-2,2)$ ms in the spike times used in (e). The red crosses and
		blue points indicate the TE values from connected and unconnected
		pairs, respectively. The black solid line indicates the threshold
		from the $k$-means method and the green dashed line indicates noise
		level. Parameters are set as $C=1\mu\textrm{F\ensuremath{\cdot}cm}^{-2}$,
		$V_{Na}=50$ mV, $V_{K}=-77$ mV, $V_{L}=-54.387$ mV, $G_{Na}=120\textrm{ mS\ensuremath{\cdot}cm}^{-2}$,
		$G_{K}=36\textrm{ mS\ensuremath{\cdot}cm}^{-2}$, and $G_{L}=0.3\textrm{ mS\ensuremath{\cdot}cm}^{-2}$,
		$V_{G}=0$ mV $\sigma_{r}=0.5$ ms, $\sigma_{d}=3.0$ ms. \cite{dayan2001theoretical,gerstner2002spiking}.
		Here the threshold $V^{\textrm{th}}=-50$ mV, $S=0.02\text{ mS/cm}^{2}$
		in (c, e, f), $\nu=0.1\textrm{ ms}^{-1}$ and $f=0.1\textrm{ mS/cm}^{2}$
		in (e, f), $k=l=1$, $\Delta t=0.5$ ms, $\tau=5$ (2.5 ms). \label{fig:HH_allin}}
\end{figure}

We now try to address the issue of whether we can use low order and
pairwise TE to reconstruct the structural connectivity of the HH network.
We start with the smallest orders $k=l=1$, $i.e.$, no memory effect.
For the network shown in Fig. \ref{fig:HH_allin}(a), we find the
TE values from connected pairs are always significantly greater than
the unconnected ones, over 100 times. This property robustly holds
when we scan $f$ and $\nu$ as shown in Fig. \ref{fig:HH_allin}(c)
which covers realistic range of firing rates 2-50 Hz. So we can simply
classify them into two groups by the $k$-means method with the larger
ones inferred from connected pairs. Then the adjacency matrix is successfully
reconstructed as shown in Fig. \ref{fig:HH_allin}(b) for all the
scanned dynamical regimes. For a network of 100 nodes randomly connected
as given in Fig. \ref{fig:HH_allin}(d), there is still a great difference
of magnitude between the TE values from the connected and unconnected
pairs as shown in Fig. \ref{fig:HH_allin}(e). By the $k$-means method,
the accuracy is 100\%. We further check whether our TE framework is
noise robust or not by adding a noise generated from the uniform distribution
$U(-2,2)$ ms in the spike times. As shown in Fig. \ref{fig:HH_allin}(f),
the great difference of magnitude still holds and the reconstruction
accuracy is 99.7\% .

We now turn to answer the question of why a low order and pairwise
TE framework can uncover the structure connectivity of the HH network
above. First we present a theoretical estimation of TE defined in
Eq.(\ref{eq:TE}). For the simplicity of writing, we map the history
state $x_{n}^{(k)}$ to a decimal number, for example $x_{n}^{(k)}=(x_{n}=1,x_{n-1}=0,x_{n-2}=0,x_{n-3}=0)$
can be one-to-one mapped to the binary number 1000 which equals to
the digital number 8. Then we can use the notations

\begin{align}
p_{a,b}(k,l,\tau) & =p(x_{n+1}=1|x_{n}^{(k)}=a,y_{n-\tau}^{(l)}=b)\\
\Delta p_{a,b}(k,l,\tau) & =p_{a,b}(k,l,\tau)-p_{a,0}(k,l,\tau)
\end{align}
where $a,b$ are the corresponding decimal numbers. We rewrite Eq.(\ref{eq:TE})
in the form of $p_{a,b}$ and $\Delta p_{a,b}$ and a Taylor expansion
gives

\begin{equation}
T_{Y\rightarrow X}(\tau)=\frac{1}{2}\sum_{a}\frac{1}{p_{a,0}-p_{a,0}^{2}}\left(\sum_{b}p(a,b)\Delta p_{a,b}^{2}\right.
\left.-\frac{(\sum_{b}p(a,b)\Delta p_{a,b})^{2}}{p(a)}\right)+O(\sum_{a,b}\Delta p_{a,b}^{3})
\label{eq:TE expression1}
\end{equation}

\begin{figure}[H]
	\begin{centering}
		\includegraphics[width=1\textwidth]{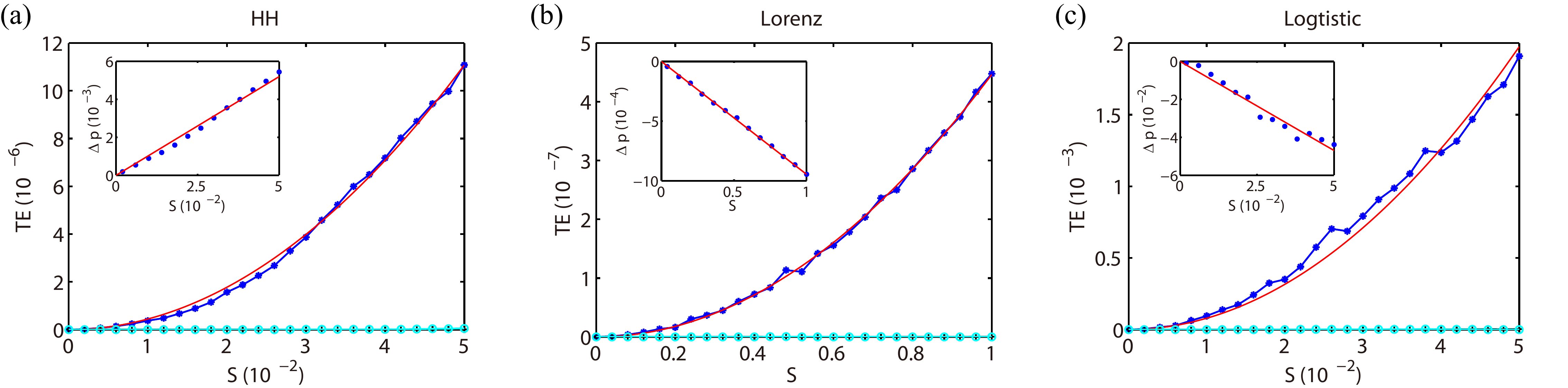}
		\par\end{centering}
	\caption{TE versus the coupling strengths of a (a) HH network, (b) Lorenz network
		and (c) Logistic network of 3 nodes connected as $Y\rightarrow X\rightarrow Z$.
		We use $T_{Y\rightarrow X}$ (blue stars), $T_{X\rightarrow Y}$ (cyan
		circles) and $T_{Y\rightarrow Z}$ (black pluses) to represent the
		direct, reverse direction and indirect TE values. Inset: linear relationship
		between $\Delta p_{0,1}$ and the coupling strengths for the connected
		pair from $X$ to $Y$. The red solid lines are quadratic fits (linear
		fits in the inset). We take $\nu=0.1\text{ms}^{-1}$ and $f=0.1$
		$\textrm{mS/cm}^{2}$ for HH network, $k=1$ for HH and Lorenz networks,
		$k=2$ for Logistic network and $l=1$ for all the networks. Time
		bins are set as $\Delta t=0.5$ ms, 0.01 and 1 for HH, Lorenz and
		Logistic networks, respectively. Delay time is set as $2.5$ ms $(\tau=5)$
		for HH network and $0$ for Lorenz and logistic networks. \label{fig:N=00003D3}}
\end{figure}

The increase $\Delta p_{a,b}$ is from the signals 1 of $Y$ and its
value is decided by the coupling strength $S$ from $Y$ to $X$.
As shown in the inset of Fig. \ref{fig:N=00003D3}, $\Delta p_{a,b}$
is linearly related to the coupling strength. We use the neuronal
system to understand the linear relationship. For the HH neurons,
an excitatory spike with strength $S=0.05\text{ mS/cm}$ can increase
the voltage of a post neuron around 1.5 mV and it requires more than
10 spikes arrived at the same time to trigger the post neuron to fire
a spike. Therefore, the influence of a single spike is very limited
and the increase $\Delta p_{a,b}$ is very small in the order of 0.01.
So the first order term is the leading order if we expand $\Delta p_{a,b}$
with respect to $S$. From Eq.(\ref{eq:TE expression1}), we can finally
conclude the relation of TE and coupling strength $T_{Y\rightarrow X}\propto S^{2}$
as shown in Fig. \ref{fig:N=00003D3}.

Due to the high dimensional problem, we apply TE under a low order
and pairwise setting. But in principle, these two setting would make
the indirect causality significant due to the memory effect and intermediate
neurons. Consider a network of 3 nodes connected as $Y\rightarrow X\rightarrow Z$.
When we measure the causality on $Y$ from $X$ by TE, the memory
effect would happen if we use a low order for $Y$. The signal of
$X$ driven by $Y$ may contain history information of $Y$ which
improves the prediction of the future of $Y$ and the computed $T_{X\rightarrow Y}$
may be significantly overestimated. However this uncertainty reduction
computed by TE is merely a self-prediction. On the other hand, when
we measure the causality on $Z$ from $Y$ without conditioning on
$X$, we may wrongly infer a significant connection from $Y$ to $Z$
which can be avoided by using a conditional TE including $X$. 

\begin{figure}[H]
	\begin{centering}
		\includegraphics[width=1\textwidth]{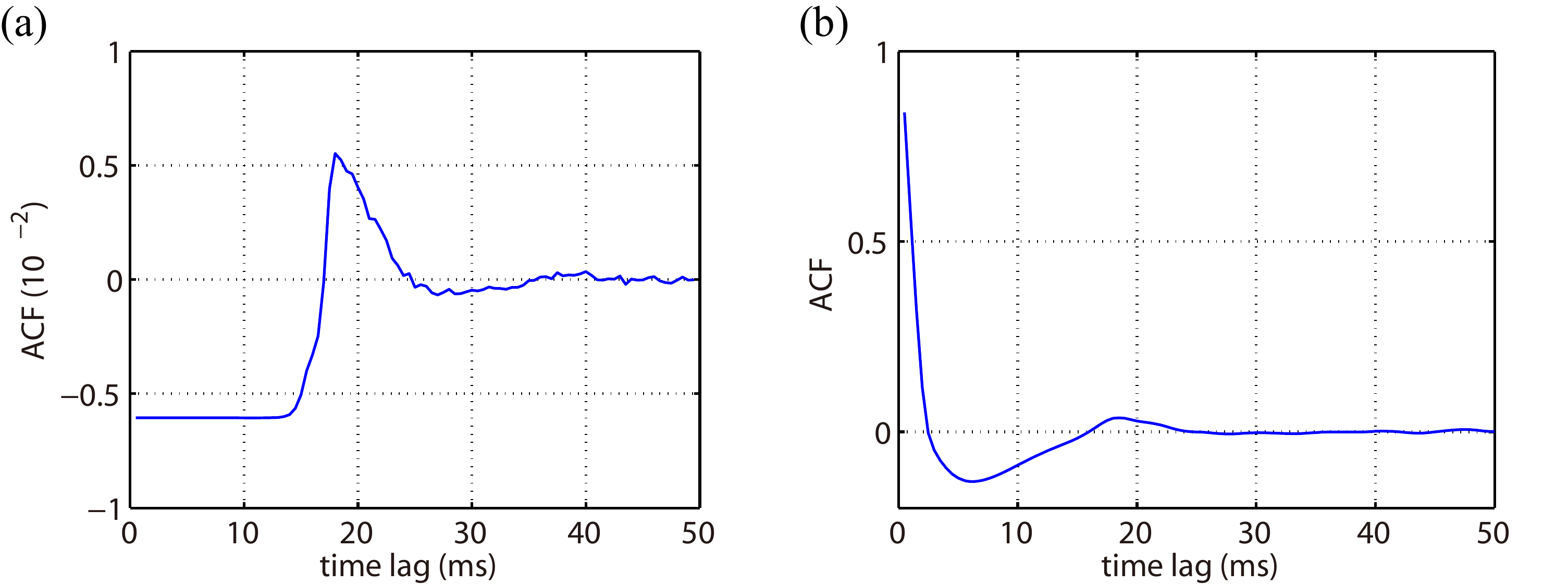}
		\par\end{centering}
	\caption{Auto-correlation function (ACF) of a HH node with binary time series
		(a) and continuous-valued voltage time series (b). Time bin $\Delta t=0.5$
		ms. \label{fig:ACF HH}}
\end{figure}

We point out that by using proper binary time series these two problems
can be avoided. For neuronal systems, the binary time series can be
directly obtained from the spike trains. Under a time bin size of
0.5 ms which resolution can be realized in experiment, we find that
almost all the components of the binary data are zeros and only a
few odd ones. The auto-correlation of the binary data is quite weak
as shown in Fig. \ref{fig:ACF HH}(a). For binary random variables,
\textcolor{black}{uncorrelatedness is equivalent to independence}.
Therefore, the obtained binary time series are almost whitened, $i.e,$
no memory effect. As shown in Fig. \ref{fig:N=00003D3}, the TE values
from reverse direction causality $T_{X\rightarrow Y}$ is negligible
even with a order of $1$. However if we use the continuous\textendash valued
voltage time series, the memory is over 10 ms as show in Fig. \ref{fig:ACF HH}(b)
and we can not use a low order. For the indirect causality, we have
$\Delta p(Y\rightarrow Z)=O(\Delta p(Y\rightarrow X)\Delta p(X\rightarrow Z))$
from the causal path. When we take a time bin size of 0.5 ms, the
increase $\Delta p$ from direct causality is quite small $O(0.01)$.
Hence the indirect $\Delta p$ is much smaller. From Eq.(\ref{eq:TE}),
the TE values from indirect causality $T_{Y\rightarrow Z}$ are also
negligible as shown in Fig. \ref{fig:N=00003D3}. 

As a model-free method, our low order and pairwise TE framework with
binary data should also work in other systems. Here we give two examples:
the Lorenz system and discrete logistic map system. The $i$th node
of a Lorenz system is governed by 
\begin{equation}
\begin{alignedat}{1}\frac{dx_{i}}{dt} & =\sigma(y_{i}-x_{i})+\sum_{j\neq i}A_{ij}\sum_{l}S\delta(t-T_{jl})\\
\frac{dy_{i}}{dt} & =\rho x_{i}-y_{i}-x_{i}z_{i}\\
\frac{dz_{i}}{dt} & =-\beta z_{i}+x_{i}y_{i}
\end{alignedat}
\end{equation}
where $\sigma=10,\beta=8/3,\rho=28$ and we take the threshold $x^{\textrm{th}}=10$.
When $x_{i}$ reaches the threshold, it will give a pulse to all the
post nodes and we denote the moment by $T_{il}$. The binary data
are obtained in the same way as in the HH system. For the logistic
map network, the $i$th node is govern by

\begin{equation}
X_{i}(n+1)=rX_{i}(n)(1-X_{i}(n))+\sum_{j\neq i}A_{ij}\sum_{l}S\delta_{i}(n+1,T_{jl})
\end{equation}

\begin{equation}
\delta_{i}(n+1,T_{jl})=\begin{cases}
1, & T_{jl}=n+1\textrm{ and }X_{i}(n+1)+S\in(0,1)\\
0, & \textrm{else}
\end{cases}\label{eq:logistic-pulse}
\end{equation}
where $r=4$ and we take the threshold $X^{\textrm{th}}=0.9$. Here
$T_{il}$ is the time when $X_{i}(T_{il}-1)<X^{\textrm{th}}$ and
$X_{i}(T_{il})\geq X^{\textrm{th}}$. For the binary data, only the
times of $\{T_{il}\}$ is 1. Note that for discrete systems, we cannot
remove the memory effect by using a small time bin but have to include
the strongly correlated historical lags which can be estimated by
the auto-correlation function as shown in Fig. \ref{fig:ACF HH}(b). 

\begin{figure}[H]
	\begin{centering}
		\includegraphics[width=1\textwidth]{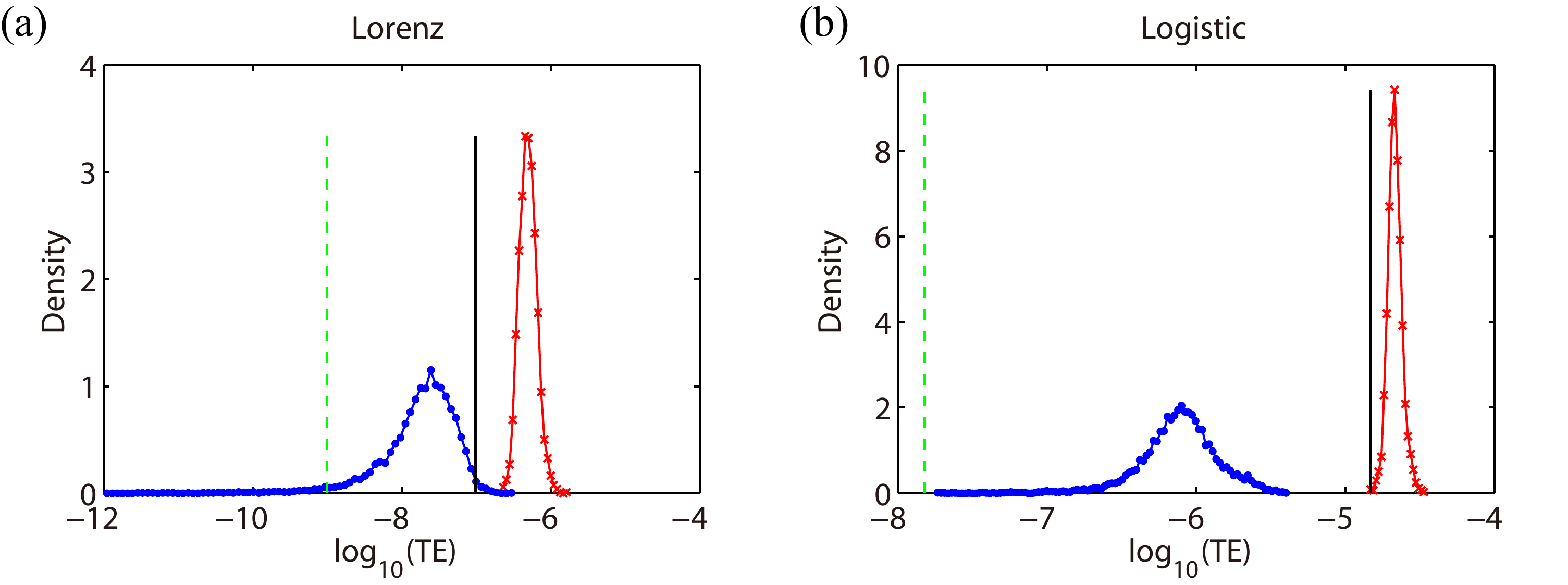}
		\par\end{centering}
	\caption{The density of $\log_{10}(\textrm{TE})$ values for (a) Lorenz network
		and (b) Logistic network. The connectivity structure and colors are
		the same as the one in Fig. \ref{fig:HH_allin}. The coupling strength
		is $S=0.5$ and $S=0.005$ for Lorenz and Logistic networks, respectively.
		Other parameters are the same as that in Fig. \ref{fig:N=00003D3}.
		\label{fig:N=00003D100}}
\end{figure}

As shown in Fig.\ref{fig:N=00003D3}, the TE framework for Lorenz
and logistic systems has similar performance as that for HH system.
Especially, the significant difference of magnitude between TE values
from connected and unconnected pairs still holds. We also apply TE
framework to Lorenz and Logistic networks of 100 nodes connected as
the one in Fig. \ref{fig:HH_allin}(d). Classified by $k$-means method,
the reconstruction accuracy is 98.7\% and 100\% for Lorenz and logistic
networks respectively as shown in Fig. \ref{fig:N=00003D100}. 

\section*{Discussion}
In summary, we have proposed a low order and pairwise TE framework
with binary data to avoid the high dimensional problem to reconstruct
the structure connectivity by detecting the TE causality among a network.
We have found and explained the phenomenon that there is a significant
difference of magnitude between TE values from connected and unconnected
pairs, depending on which the structure connectivity can be easily
reconstructed by cluster methods. Our TE framework can be applied
to a wide class of systems like the non-linear, discrete-valued or
continuous-valued ones with a high reconstruction accuracy. We have
also established a quadratic relationship between the TE values and
the coupling strengths. 

We should first point out that our TE framework does not rely on pulse-coupled
dynamics as we used above. For example, we can use a continuous function
to describe the dynamics of synaptic interactions in the HH model
\cite{compte2003cellular,sun2010pseudo}

\begin{equation}
\frac{dH_{i}(t)}{dt}=-\frac{H_{i}(t)}{\sigma_{d}}+f\sum_{l}\delta(t-T_{il}^{F})+\sum_{j\neq i}A_{ij}\sum_{l}Sg(V_{j}^{\textrm{pre}})\label{eq:HH dH2}
\end{equation}
where 
\begin{equation}
g(V_{j}^{\textrm{pre}})=\frac{1}{1+\exp\left(-\left(V_{j}^{\textrm{pre}}-20\right)/2\right)}\label{eq:HH dH3}
\end{equation}
We can also extend the Lorenz system to a continuously coupled case
\cite{martini2011inferring,belykh2004connection,belykh2006synchronization}
\begin{equation}
\frac{dx_{i}}{dt}=\sigma(y_{i}-x_{i})+\sum_{j\neq i}A_{ij}S(x_{j}-x_{i})
\end{equation}
and choose a proper threshold to obtain suitable binary time series.
Conclusions shown in Figs. \ref{fig:N=00003D3} and \ref{fig:N=00003D100}
will not change in these two extended models.

As for the proper time delay, it depends on the detailed systems.
For example, once a recipient HH node receives a spike from a driver
HH node, its voltage will increase and reach a local peak value some
time later, around which is the optimal time delay. Another way is
to scan the time delay to reach a peak TE value. In our TE framework,
the order of magnitude makes sense, so the proper time delay allows
a relatively wide range. 

Finally, we point out that the distributions of TE values from connected
and unconnected pairs shown in Figs. \ref{fig:HH_allin} and \ref{fig:N=00003D100}
may have a large overlap, $e.g.,$ when the coupling strength is inhomogenous
and follows a distribution. Then the $k$-means method may wrongly
predict the connected pairs with weak couplings and unconnected pairs
with strong indirect causality. A future work that may partially revise
this problem is that we first preliminary compute the pairwise TE
values and obtain the inferred adjacency matrix, then for each pair
we recompute TE conditioned on the important intermediate nodes obtained
from the preliminary inferred adjacency matrix and classify them by
$k$-means method again.

\section*{Acknowledgments}
This work was supported by NYU Abu Dhabi Institute G1301 (Z.K.T, D.Z., and D.C.), NSFC-11671259, NSFC-11722107, NSFC-91630208, Shanghai Rising-Star Program-15QA1402600 (D.Z.), NSFC 31571071, NSF DMS-1009575 (D.C.), Shanghai 14JC1403800, Shanghai 15JC1400104, SJTU-UM Collaborative Research Program (D.Z. and D.C.).\\

\begin{small}
\bibliographystyle{plain}
\bibliography{TERefer}
\end{small}

\end{document}